\documentclass[12pt]{iopart}
\usepackage[utf8]{inputenc}
\usepackage{iopams}  

\usepackage{graphicx}
\usepackage{dcolumn}
\usepackage{bm}

\usepackage{braket}

\usepackage{color,soul}
\usepackage{wasysym}
\usepackage{mathrsfs}
\usepackage{times}

\usepackage[dvipsnames,svgnames,table]{xcolor}
\usepackage{hyperref}
\usepackage{fancyhdr}
\usepackage{float}
\usepackage[english]{babel}
\usepackage[caption=false]{subfig}

\newcommand{\bk}{{\bf k}}

\begin{document}

\title[Perspective on topological states of non-Hermitian lattices]{Perspective on topological states of non-Hermitian lattices}

\author{Luis E. F. Foa Torres}%$^*$
%\email{lfoa@ing.uchile.cl}

\address{Departamento de F\'{\i}sica, Facultad de Ciencias F\'{\i}sicas y Matem\'aticas, Universidad de Chile, Santiago, Chile}
\ead{luis.foatorres@uchile.cl}

\begin{abstract}
The search of topological states in non-Hermitian systems has gained a strong momentum over the last two years climbing to the level of an emergent research front. In this Perspective we give an overview with a focus in connecting this topic to others like Floquet systems. Furthermore, using a simple scattering picture we discuss an interpretation of concepts like the Hamiltonian's defectiveness, i.e. the lack of a full basis of eigenstates, crucial in many discussions of topological phases of non-Hermitian Hamiltonians.

\end{abstract}

%\maketitle

\section{Introduction}

More than one hundred years after Edwin Hall discovered the Hall effect, the finding of the integer quantum Hall effect~\cite{von_klitzing_new_1980} and its topological origin~\cite{thouless_quantized_1982} was followed by a sequence of breakthroughs~\cite{haldane_model_1988,kane_quantum_2005,bernevig_quantum_2006} that sprouted into the discovery of topological insulators in two~\cite{konig_quantum_2007} and three dimensions~\cite{hsieh_topological_2008}. Unlike conventional phases of matter, the new topological phases~\cite{von_klitzing_new_1980,thouless_quantized_1982,haldane_model_1988,kane_quantum_2005,bernevig_quantum_2006,konig_quantum_2007,hsieh_topological_2008} cannot be described by a local order parameter~\cite{bernevig_topological_2013,frank_ortmann_topological_2015,asboth_short_2016}. Now, the frontiers in this field keep expanding at a rapid pace in ever more interesting directions, like the search for gapless but topological phases such as Weyl semimetals~\cite{vafek_dirac_2014,hasan_discovery_2017} and topological states in Floquet systems~\cite{oka_photovoltaic_2009,lindner_floquet_2011,mciver_light-induced_2018}.

One of the latest research fronts in this area is the search for topological states of non-Hermitian lattices~\cite{rudner_topological_2009,diehl_topology_2011,yuce_topological_2015,san-jose_majorana_2016,lee_anomalous_2016,leykam_edge_2017}. The interest on non-Hermitian Hamiltonians was originally focused in $\mathcal{P T}$-symmetric Hamiltonians~\cite{bender_real_1998} as a generalization of quantum mechanics where the Hermiticity constraint could be removed while keeping a real spectra. Today, this has shifted to non-Hermitian Hamiltonians regarded as an effective description of, for example, open quantum systems~\cite{rotter_non-hermitian_2009,rotter_review_2015},  where the finite lifetime introduced by electron-electron or electron-phonon interactions~\cite{kozii_non-hermitian_2017,yoshida_non-hermitian_2018}, or disorder~\cite{zyuzin_flat_2018}, is modeled through a non-Hermitian term, or in the physics of lasing~\cite{liertzer_pump-induced_2012,peng_chiral_2016,harari_topological_2018,bandres_topological_2018,zhao_topological_2018}. An additional source of momentum in this field comes from the study of systems where the quantum mechanical description is used after mapping to a Schr\"odinger-like equation, as in systems with gain and loss (as found in optics and photonics~\cite{ruter_observation_2010,longhi_parity-time_2017,feng_non-hermitian_2017,ozawa_topological_2019}), surface Maxwell waves~\cite{bliokh_topological_2019}, and topoelectrical circuits~\cite{lee_topolectrical_2018,ezawa_non-hermitian_2019}.

Topological effects are intrinsic to features unique to non-Hermitian Hamiltonians such as the topological structure of exceptional points, points in parameter space where the eigenvalues and eigenvectors coalesce (the non-Hermitian counterpart of Hermitian degeneracies), and have been studied for a few decades 
~\cite{dembowski_experimental_2001,dembowski_encircling_2004,stehmann_observation_2004,hu_exceptional_2017,miri_exceptional_2019}. The focus of the present new wave of interest is, however, on a different aspect: given a non-Hermitian lattice, a system where a \textit{motif} is periodically repeated in space, what is the phenomenology and how do we classify the topological nature of the resulting states? Here by topological one means that there is underlying bulk-boundary correspondence relating a bulk topological invariant to the existence (or absence) of states localized at a boundary (between topologically different phases). 

The search of topological states of non-Hermitian lattices has become one of the most exciting emergent fronts at the crossroads of condensed matter physics, optics and photonics, acoustics and quantum physics. Together with Floquet systems, non-Hermitian lattices are our first strong feet on the land of non-equilibrium, dynamical, topological phases. The departure from the Hermitian paradigm brings many surprises, like the fact that because of the extreme sensitivity to boundary conditions, a pristine non-Hermitian lattice may become devoid of extended states~\cite{xiong_why_2018,martinez_alvarez_non-hermitian_2018}, an effect termed non-Hermitian skin effect~\cite{yao_edge_2018,martinez_alvarez_topological_2018}.
On the other hand, there is an intense activity in the search of a consistent classification~\cite{shen_topological_2018,gong_topological_2018,yin_geometrical_2018,kawabata_symmetry_2018,li_geometric_2019}, the definition of the topological invariants~\cite{yao_non-hermitian_2018,ghatak_new_2019,jiang_topological_2018}, and the search of a bulk-boundary correspondence~\cite{xiong_why_2018,martinez_alvarez_non-hermitian_2018,kunst_biorthogonal_2018,borgnia_non-hermitian_2019}.

Here we provide an overview of this fascinating topic. We start by discussing how to define a gap for a complex spectrum. Later on, we discuss the information encoded in the richer spectrum of non-Hermitian systems and how to make sense of the non-Hermitian terms. In Section 5, we discuss non-Hermitian degeneracies (exceptional points) and later on in Section 6, by resorting to a scattering picture, we put forward an interpretation of the Hamiltonian defectiveness (the lack of a full basis of eigenvectors). In Section 7 we give a flavor of the many paths proposed for building a bulk-boundary correspondence. Finally, in Section 8, we argue on the connections with Floquet systems.

\section{Defining a gap for a complex spectrum}
\label{section-gap}

Unlike Hermitian Hamiltonians, non-Hermitian systems have eigenvalues with a real and an imaginary part~\cite{moiseyev_non-hermitian_2011}. Thus, since complex numbers do not have an order relation, we cannot recur to the usual (Hermitian) definition of a gap~\cite{martinez_alvarez_topological_2018}. 

To fix ideas let us consider a non-Hermitian lattice with Bloch-type eigenstates and eigenenergies $E_n(\bk)$, where $\hbar \bk $ is the quasimomentum and $n$ is the band index. 
There are many possible extensions of the gap concept. Ref.~\cite{shen_topological_2018} defines a band (say $ n $) as \textit{separable} if $E_n(\bk) \neq E_m(\bk)$ for all $m \neq n$ and all $\bk$;  \textit{isolated} if $E_n(\bk) \neq E_m(\bk')$ for all $m \neq n$ and all $\bk, \bk'$; \textit{inseparable} when for some quasimomentum the (complex) energy becomes degenerate with that of another band.

Alternatively one can generalize the concept of a bandgap as \textit{the prohibition of touching a base energy}~\cite{gong_topological_2018}, also called \textit{point gap}. This base energy is most typically set to zero but it can be complex in the general case. In the Hermitian case we can continuously deform the spectrum in such a way that each energy band is mapped to a point along the (real) energy axis, all without touching a base energy set inside the gap (and therefore closing it). In the non-Hermitian case, when we sweep over the Brillouin zone the bands describe loops in the complex plane. If the base point lies inside the loop, the loop cannot be deformed onto a point without touching the base.

\begin{figure}
\centering
\includegraphics[width=0.60\linewidth]{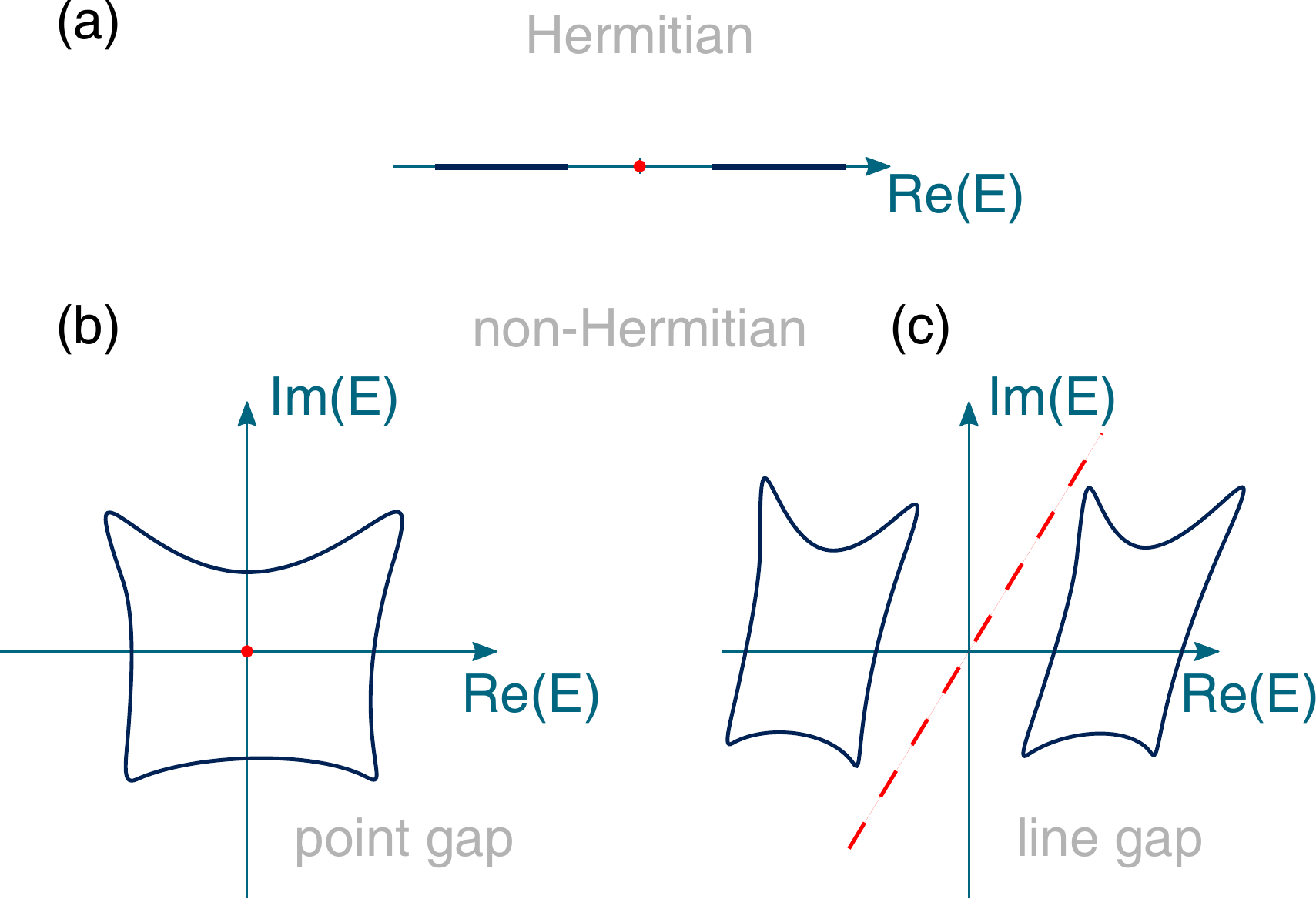}
\caption{(a) Scheme representing the eigenenergies forming bands (thick segments) in a typical gapped Hermitian system. The red dot marks the Fermi energy. In the non-Hermitian case, the gap can be defined as the prohibition of touching a base energy, also called point gap (b), or the prohibition of touching a line, also called line gap (c).}
\label{fig1}
\end{figure}

Thus, the prohibition of touching a base energy~\cite{gong_topological_2018} is one physically reasonable generalization of the gap concept. Having a spectrum which describes a loop in a two-dimensional space allows to define a topological invariant as the winding number of that loop, without any reference to the eigenstates~\footnote{Note that this is in stark contrast with the topological invariants in hermitian systems, which depend only on the eigenstates and not the eigenenergies.}. This winding number appears in many proposals for classifying non-hermitian lattices including that by Gong et al.~\cite{gong_topological_2018}, Shen and Fu~\cite{shen_topological_2018} (where it is dubbed \textit{vorticity}). But the reader might have realized that this opens up the possibility of having a non-trivial topological invariant even in a system with a \textit{single} band! Indeed, a single loop (and hence a single band) enclosing the base energy point gives a non-vanishing winding number. This contrasts with the Hermitian case, where at least two bands are required to have a non-zero Chern number \footnote{Recall that the sum of the Berry curvature over all available bands is zero at each point in parameter space.}. Here, having a different invariant allows for the rules to change.

Finally, one can also define the \textit{line gap}: If the energy bands do not cross a line in the complex energy plane, the system is said to have a line-gap~\cite{kawabata_symmetry_2018,borgnia_non-hermitian_2019}. This concept accommodates the idea of a secluded region in the complex energy plane between two bands. The point and line gap definitions are complementary in the sense that one might be more relevant than the other in a particular situation. The existence of this two types of gaps is also a consequence of moving from a one-dimensional energy line for Hermitian systems to a plane for the case with a complex spectrum: In the Hermitian case, the secluded regions of energy can only be zero-dimensional (e.g. a segment of the line as in Fig.~\ref{fig1}(a)), while in the non-Hermitian one can either be zero-dimensional (point gap, as in Fig.~\ref{fig1}(b)) or one-dimensional (line gap, as in Fig.~\ref{fig1}(c)).

\section{The intriguing imaginary part}

Up to now we have seen how taking the energy spectrum to the complex plane brings some surprises. But what does it mean to have an imaginary part of the eigenergies and what information does it encode? A very first point is that non-Hermitian systems can be regarded as non-equilibrium systems. From a dynamical viewpoint, the states may even be unstable as the time-dependent Schr\"odinger equation is non-unitary. For long enough times, only the state(s) with the largest imaginary part of the eigenenergy survives.

Non-hermiticity however, does not necessarily lead to a complex spectrum~\cite{bender_real_1998,martinez_alvarez_topological_2018}. If a Hamiltonian has $\mathcal{P T}$ symmetry~\footnote{$\mathcal{H}$ is said to be $\mathcal{P T}$-symmetric if the combined action of the parity $\mathcal{P}$ and time-reversal $\mathcal{T}$ operators leave it invariant, $\mathcal{H}=(\mathcal{P T})^{\dag} \mathcal{H}(\mathcal{P T})^{ }$.} and that symmetry is  preserved by the eigenstates then the corresponding eigenvalues are real~\cite{bender_real_1998}.
%Even when for a PT-symmetric Hamiltonian one may have a region of states preserving PT-symmetry with real eigenvalues
But even in that case, the vanishing imaginary part stems from a dynamical situation where gains and losses are compensated.

In spite of the problems brought by non-Hermiticity, the imaginary part of the spectrum brings opportunities to clearly define things that prove difficult within standard quantum mechanics as, for example, time scales. Being not a dynamical variable but rather a parameter, time does not have a corresponding operator in quantum mechanics, thereby complicating the determination of time scales such as the tunneling time~\cite{landauer_barrier_1994}. Many authors (see~\cite{rotter_non-hermitian_2009,rotter_review_2015} and references therein) have pointed out that from a non-Hermitian perspective, the imaginary part of the energies provide precisely those evasive time scales, almost for free.

\section{Making sense of gains, losses and asymmetric couplings} 

The conventional Hermitian quantum mechanics is so strongly built in our thinking that to many of us it might seem devoid of physical meaning to speak about non-Hermitian Hamiltonians. However, non-Hermitian Hamiltonians arise as a natural effective description of different systems. This includes photonic~\cite{ruter_observation_2010,longhi_parity-time_2017,ozawa_topological_2019} and acoustic systems where gains and losses are naturally present, and also open quantum systems. In the latter case, non-Hermiticity arises when tracing out over the infinite degrees of freedom of what we consider as the \textit{environment} acting on \textit{the sample} of interest. Although defining the line that separates what belongs to the sample and what remains in the environment is arbitrary, the response obtained, for example through Green function techniques, is exact and independent of where we decided to draw this line.

The reader at this point might be puzzled about the fact that the effective Hamiltonians typically encountered in electronic systems contain only losses, because the eventual leak into the environment, but no gains. How to make sense of gains when all the physical properties seem to be well described by the retarded self-energy corrections? The answer is in the boundary conditions. The information concerning the boundary conditions in a scattering problem, e.g. the particle is coming from the left or right, is absent in the effective Hamiltonian containing usual retarded self-energy corrections. This information has to be supplemented somehow if we want to compute the scattering state projected on the sample. One possible path is the Lippmann-Schwinger equation, another is akin to flipping the sign of the imaginary part of the self energy correction to signal where the source is, the latter provides a non-Hermitian Hamiltonian with losses \textit{and} gains.

In most models studied in the context of topological states, gains and losses might be distributed over the system, thus allowing for a lattice structure, or they might have some symmetry with respect to an interface, thereby forming a boundary between two regions~\cite{ni_$mathcalpt$_2018,yuce_edge_2018,ryu_emergent_2019}.

Besides gains and losses, a non-Hermitian Hamiltonian can also have asymmetric couplings between states, i.e. $\bra{i}\mathcal{H}\ket{j} ^{} \neq \bra{j}\mathcal{H}\ket{i} ^{*}$. This sort of asymmetric or non-reciprocal hoppings appear in many models~\cite{yao_non-hermitian_2018,lee_anatomy_2019,lee_hybrid_2019}, and though they can be more difficult to assimilate in condensed matter systems (unless one thinks of them as an effective model or just a mapping between a model with gain and losses and another one with nonreciprocal hoppings) they are common in photonics.

\section{Exceptional points and defectiveness}

Having a non-Hermitian Hamiltonian ${\cal H}={\cal H}_0+\lambda {\cal H}'$ does not lead by itself to new physics, for example, if the non-Hermitian term ${\cal H}'$ commutes with the Hermitian ${\cal H}_0$, the eigenstates do not change. One of the key features unique to non-Hermitian Hamiltonians is defectiveness. This is, the eigenvectors of a non-Hermitian matrix can have an incomplete set of eigenvectors. Take for example the simplest Jordan block, a $2\times 2$ matrix with all elements equal to $1$ except for the lower left element which is set to zero. It has the eigenvalue $\lambda=1$ with degeneracy two but has a single eigenvector. This matrix is said to be \textit{defective}, a situation that is truly unique to non-Hermitian Hamiltonians.

\begin{figure}
\centering
\includegraphics[width=0.75\linewidth]{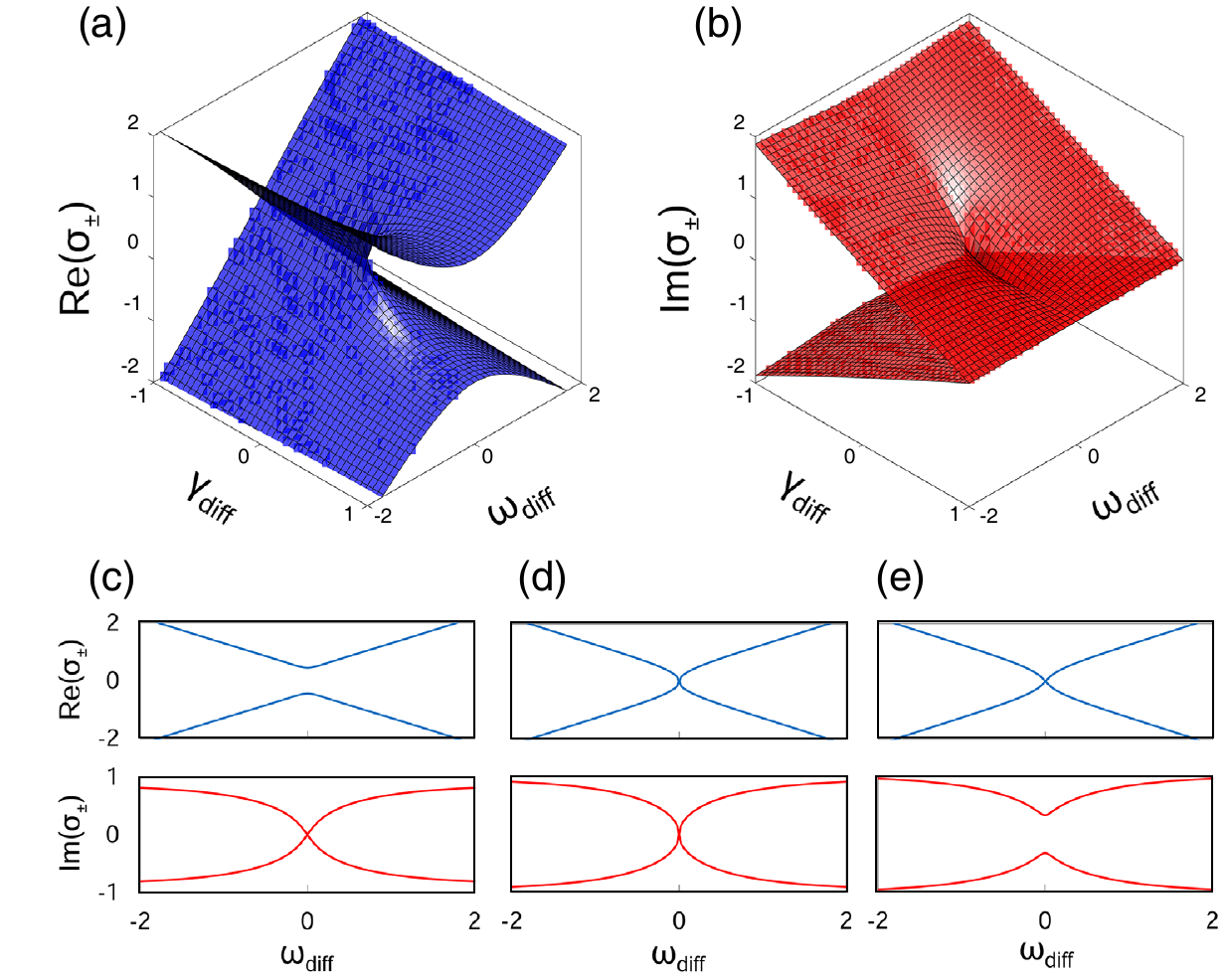}
\caption{ $2 \times 2$ Hamiltonian with complex onsite terms $\omega_{\mathrm{diff}}-i \gamma_{\mathrm{diff}}$ and $-\omega_{\mathrm{diff}}+i \gamma_{\mathrm{diff}}$ and reciprocal hopping $v$ exhibiting an exceptional point. (a) and (b) show the evolution of the real and imaginary parts of the eigenvalues, denoted with $\sigma_{\pm}$ as a function of the difference in frequencies $\omega_{\mathrm{diff}}$ and loss factors $\gamma_{\mathrm{diff}}$.  (c-e) show cuts for different values of $\gamma_{\mathrm{diff}}$: for $\gamma_{\mathrm{diff}}>0$ (c) there is level repulsion in the real part of the eigenvalues, for $\gamma_{\mathrm{diff}}=0$ (d) the real and imaginary parts coalesce, while for $\gamma_{\mathrm{diff}}<0$ the real parts cross and the imaginary ones repel each other. From Miri and Al\`u, Science \textbf{363}, eaar7709 (2019)~\cite{miri_exceptional_2019}. Reprinted with permission from AAAS.}
\label{fig2}
\end{figure}

Defectiveness typically emerges at the so-called exceptional points. Let us think of a generic situation with a Hamiltonian of the form ${\cal H}={\cal H}_0+\lambda {\cal H}_1$, where $\lambda$ is a parameter controlling the strength of the second term. Assuming Hermitian ${\cal H}_0$ and ${\cal H}_1$, the levels repel each other and avoid crossing when $\lambda$ changes~\cite{von_neuman_uber_1929}, as long as $\lambda$ is real. But when $\lambda$ is allowed to adopt complex values, the levels can \textit{coalesce}, i.e. they can merge while rendering the matrix defective. This coalescence appears as a square root singularity of the eigenenergies as a function of the parameter $\lambda$ and is called an exceptional point~\cite{kato_perturbation_1995} (for a recent review on these points in photonics see~\cite{miri_exceptional_2019}). The importance of these points was emphasized by Berry ~\cite{berry_pancharatnam_1994,berry_geometric_2010}. Because of the complex shape of the energy Riemann surfaces in the vicinity of an exceptional point (see Fig.~\ref{fig2}), which also lead to a breakdown of the adiabatic theorem, there are many counter-intuitive properties such as a strong dependence on the parameter $\lambda$ close to those points, or the phenomenon dubbed chiral state conversion~\cite{hassan_chiral_2017,hassan_erratum:_2017,zhang_dynamically_2018}. \footnote{We note that there has been an intense recent discussion on the need (or lack of it) to encircle the exceptional point to obtain chiral state conversion~\cite{hassan_chiral_2017,hassan_erratum:_2017} and also on the dependence on the initial point~\cite{zhang_dynamically_2018}. We urge the reader to get updated on these issues by reading those references.} This has fueled fascinating experiments with microwaves ~\cite{dembowski_experimental_2001,dembowski_encircling_2004,stehmann_observation_2004,bittner_$mathscpmathsct$_2012,hu_exceptional_2017,doppler_dynamically_2016,zhang_dynamically_2018}, optical waveguides~\cite{ghosh_exceptional_2016}, nuclear magnetic resonance~\cite{alvarez_environmentally_2006}, and also in optomechanical systems~\cite{xu_topological_2016}. Exceptional points lead to intriguing phenomena such as unidirectional invisibility~\cite{lin_unidirectional_2011}, single-mode lasers~\cite{feng_single-mode_2014,hodaei_parity-timesymmetric_2014}, or enhanced sensitivity in optics~\cite{chen_exceptional_2017,hodaei_enhanced_2017,rechtsman_applied_2017}. Many of these phenomena could be understood in terms of an environment mediated interaction~\cite{alvarez_environmentally_2006,bird_electron_2004,yoon_coupling_2012,rotter_review_2015,auerbach_super-radiant_2011,eleuch_resonances_2017}.

Exceptional points may also acquire different flavors by becoming anisotropic~\cite{xiao_anisotropic_2019}, packing onto exceptional surfaces as reported by Okugawa and Yokoyama~\cite{okugawa_topological_2019}, exceptional lines driven by disorder~\cite{moors_disorder-driven_2019}, or exceptional rings~\cite{yoshida_exceptional_2019, yoshida_symmetry-protected_2019}. They can also be engineered~\cite{molina_surface_2018}
or used as a means to interpret the physics of surface states in three-dimensional systems ~\cite{gonzalez_topological_2017}.

For $\mathcal{P T}$ symmetric Hamiltonians, exceptional points mark the onset of a transition from a region with real eigenvalues (and thus oscillatory solutions) to one with complex (conjugate) eigenvalues (and thus unstable growing or decaying solutions), all in spite of the evolution respecting always $\mathcal{P T}$ symmetry (this corresponds to a cut along $\gamma_{\mathrm{diff}}$ in Fig.~\ref{fig2} (a-b))~\footnote{We note that the transition from $\mathcal{P T}$ preserved to $\mathcal{P T}$ broken eigenstates may be shifted away from the exceptional points due to non-linearities as reported in ~\cite{ge_anomalous_2016}.}.

\section{Interpreting defectiveness}

\begin{figure}
\centering
\includegraphics[width=0.85\linewidth]{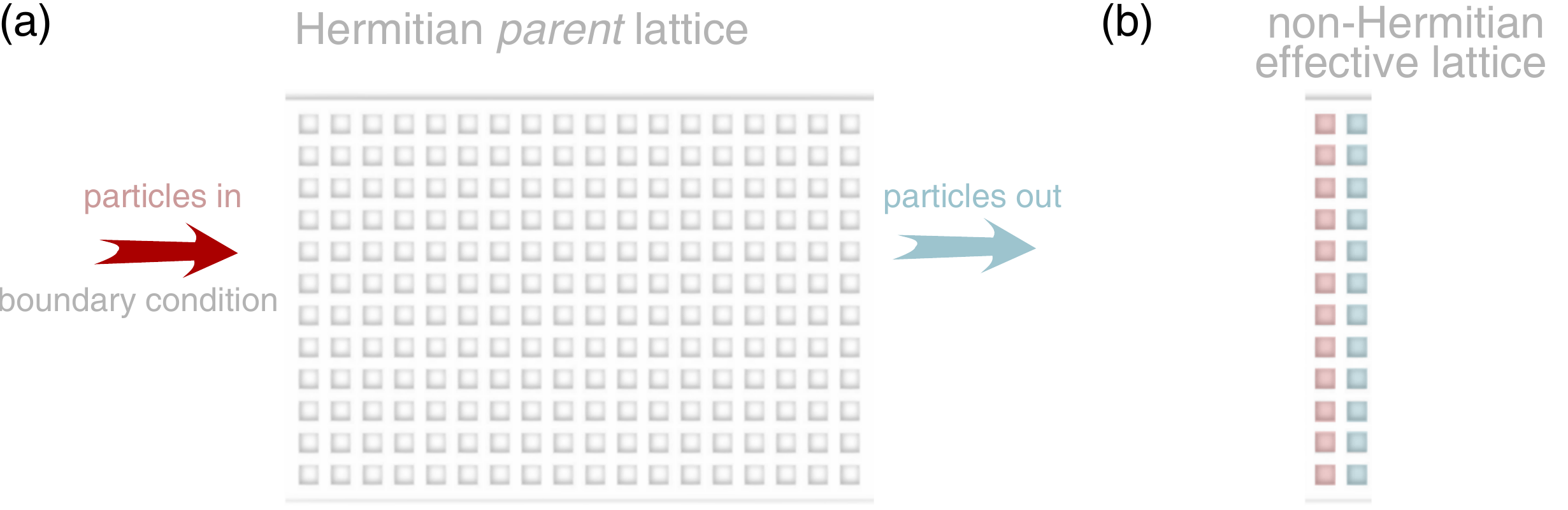}
\caption{(a) Scheme representing a hypothetical Hermitian lattice with translational invariance along the horizontal direction. The red arrow represents the (boundary) condition of particles incoming from the left. (b) When projected on a strip the situation in (a) can also be described by an effective non-Hermitian Hamiltonian (gains and losses stemming from the boundary condition are represented in color).}
\label{fig3}
\end{figure}

In spite of the pivotal role of defectiveness in phenomena unique to non-Hermitian systems, no interpretation is usually given in the context of the search for topological states. Here, we use the simple scattering picture mentioned in previous sections to make sense of defectiveness. The main idea is that a lattice with at least one translationally invariant direction (in $d-$spatial dimensions) under a non-equilibrium or dynamic condition where, for example, the particles are bounded to come from the left (e.g. a boundary condition) can be mapped to a non-Hermitian lattice in dimension $d-1$ spanning a portion of the original Hilbert space, this is represented in Fig.~\ref{fig3}. The effective Hamiltonian for the reduced system is such as to provide the scattering states of the ancestor or parent Hermitian lattice projected onto the chosen Hilbert space. This path has been explored earlier in tight-binding lattices while trying to make sense of non-Hermitian Hamiltonians in a series of studies by Jin and Song~\cite{Jin_physics_2010,jin_physical_2011} and others~\cite{hernandez-coronado_perfect_2011}.

We start by mentioning that this approach has an immediate weakness: there is no warranty that a non-hermitian Hamiltonian can always be assimilated to the effective description of a scattering situation (typically stemming from a Hermitian Hamiltonian in larger Hilbert space plus a boundary condition, e.g. incidence direction of the particles)\footnote{Furthermore, in many cases, even when this parent or ancestor Hermitian Hamiltonian exists, the effective description usually involves an energy dependent Hamiltonian (and so the link is restricted to a discrete set of energies (resonances)).}. That said, a key point is the realization that the effective Hamiltonian with gains and losses also encodes information on the boundary condition (which is typically enforced separately!). As such, defectiveness, the lack of a full set of linearly independent eigenvectors, can be understood as the lack of enough states which are compatible with the boundary condition. 

As an example of the above arguments, let us think of a ribbon made of a Chern insulator with particles incident from the left (our boundary condition as typically used to compute transmission coefficients). Now we decide that we are interested only on the projection of those scattering states compatible with the boundary condition over a stripe perpendicular to the translationally invariant direction. In this case, the effective one-dimensional system will sustain an edge state at a single edge and not both, thus reducing the number of linearly independent eigenvector(s). 

%A byproduct of the scattering picture is a dimensionality link: 
Based on the scattering picture it is thus reasonable to expect that a non-Hermitian lattice in dimension $d$ is an effective description of an Hermitian lattice in dimension $d+1$ plus boundary conditions, when such parent lattice exists. This \textit{dimensionality link} is consistent with several findings in the literature including: the observation in ~\cite{gong_topological_2018} that a simple one-dimensional chain with asymmetric hoppings is the non-hermitian proxy of the Chern insulator (as a topological phase not requiring a symmetry); the anomalous localization transitions in non-Hermitian systems found in low dimensions~\cite{hatano_localization_1996}. In this sense, the non-Hermitian effective description is very much like a shadow (in the sense of a projection) of the higher-dimensional Hermitian ancestor, in a situation reminiscent of what happens in quasicrystals~\cite{kraus_topological_2012,martinez_alvarez_edge_2019}.

\subsection{A path to obtain a Hermitian parent of a non-Hermitian Hamiltonian}

At this point the reader might wonder how to proceed to obtain the Hermitian parent of a given non-Hermitian Hamiltonian. Here we will sketch the procedure for a simple case following Jin and Song~\cite{Jin_physics_2010,jin_physical_2011}. A first point is to refine our starting question. Specifically, we want to obtain a Hermitian parent of a given non-Hermitian Hamiltonian so that a subset of eigenstates of both Hamiltonians share the same projection of the corresponding eigenstates over the common part of the Hilbert space.

\begin{figure}
\centering
\includegraphics[width=0.70\linewidth]{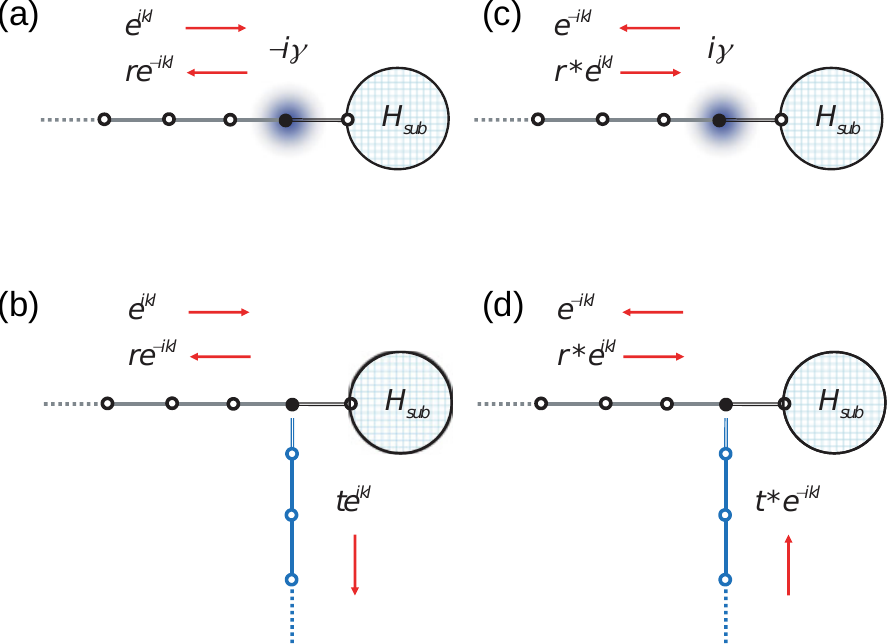}
\caption{(a) Scheme representing a tight-binding network with an on-site imaginary potential $-i\gamma$ at one site and an arbitrary network with an Hermitian Hamiltonian represented by $H_{\mathrm{sub}}$. (b) represents the Hermitian parent of the system in (a) sharing the same eigenstate within the common region. (c) shows the same system as (a) except that the sign of the imaginary term has been changed. In this case, the parent Hermitian Hamiltonian is the same as before, shown in (d), what changes is the boundary condition for incident waves (red arrows). Reprinted figure with permission from L. Jin and Z. Song, Phys. Rev. A \textbf{81}, 032109 (2010). Copyright (2010) by the American Physical Society.}
\label{fig4}
\end{figure}

In many cases an answer to the previous question can be obtained by recurring to a scattering picture. Fig.~\ref{fig4} (a) represents a tight-binding network containing containing an absorbing on-site term ($-i\gamma$) at one site. $H_{\mathrm{sub}}$ is an arbitrary network with Hermitian terms. We know that such non-Hermitian term can serve as an effective description of the interaction with the outer world, see for example~\cite{damato_conductance_1990} where this Hamiltonian approach was put forward in the context of quantum transport in the presence of decoherence effects. Therefore one can seek a Hermitian parent in the network of Fig.~\ref{fig4} (b), which is the same network as the one in (a) with the non-Hermitian term replaced by the semi-infinite lead acting as an outer world. 

To connect both systems one needs to impose that the solutions projected over the common Hilbert space coincide. The solutions can be chosen as superposition of plane waves with unknown amplitudes as indicated in the figure. Replacing this ansatz into the Schr\"odinger equation will lead to equations that need to be fulfilled so that the requirement is met. Jin and Song~\cite{Jin_physics_2010} showed that the conditions lead to an interesting connection: real-energy eigenstates of PT-symmetric non-Hermitian tight-binding Hamiltonians have a counterpart (with equal projection over the common Hilbert space): the resonant transmission states of the Hermitian lattice~\footnote{Further work by the same authors generalized this result to situation where the PT-symmetry requirement is relaxed~\cite{jin_physical_2011}.}.

If we consider the same system but now invert the sign of the imaginary on-site potential (thus representing a source rather than a sink) as shown in Fig.~\ref{fig4} (c), one can show that the connection with the Hermitian parent Hamiltonian also occurs but this time at a different state: a resonant state (shown in Fig.~\ref{fig4} (d)) which is the time-reversed solution of the one in Fig.~\ref{fig4} (b). Therefore, we see that the full Hermitian parent Hamiltonian has two linearly independent solutions related by TRS, but each of them corresponds to a \textit{different} non-Hermitian child and each of them is defective at that energy. We can see that the boundary condition on the Hermitian parent necessary to define the scattering states, affects the non-Hermitian child.

\subsection{Extreme defectiveness from higher-order exceptional points and the non-Hermitian skin effect}

Finally, we comment on an extreme case of defectiveness found in non-Hermitian lattices: A pristine (without disorder) finite system may become devoid of extended states having all eigenstates localized close to a boundary~\cite{xiong_why_2018,martinez_alvarez_non-hermitian_2018}, the same systems typically show an extreme sensitivity of the spectrum to a change in the boundary conditions (from open to periodic)~\cite{xiong_why_2018}. This anomalous localization was a attributed to the proximity (in parameter space) to exceptional points with an order scaling with the system size in Ref.~\cite{martinez_alvarez_non-hermitian_2018}, an effect which was interpreted as an environment mediated interaction effect. The same effect was confirmed analytically in Ref.~\cite{yao_edge_2018} for a non-Hermitian SSH model with asymmetric hoppings and dubbed \textit{non-Hermitian skin effect}. Recently, Lee and Thomale~\cite{lee_anatomy_2019} presented a characterization of these peculiar boundary modes and provided conditions for their existence. The non-Hermitian skin effect puts a conceptual obstacle for building a bulk-boundary correspondence as typically the topological invariants are based on the bulk eigenstates of the system, which in these cases may have a completely different character that the eigenstates of the finite system, no matter how large~\cite{xiong_why_2018}.

\section{The many paths to a bulk-boundary correspondence}

The term topological is used in connection to a given state to mean that there is an underlying bulk-boundary correspondence. Establishing a bulk-boundary correspondence in the most usual sense requires: (\textit{i}) characterizing the bulk states through a topological invariant, typically an integral over the the Brillouin zone of a kernel that depends only on the bulk (Bloch-type) eigenstates; (\textit{ii}) characterizing the edge or boundary states when translational symmetry is lost in at least one space direction; (\textit{iii}) linking the topological invariant with the boundary modes. Since the pioneering TKNN paper~\cite{thouless_quantized_1982}, this has been a crucial issue where progress is still underway even in Hermitian Hamiltonians~\cite{rhim_bulk-boundary_2017,rhim_unified_2018}, the interested reader can find more extensive accounts on this topic in Refs.~\cite{bernevig_topological_2013,asboth_short_2016,ortmann_topological_2015,vanderbilt_berry_2018}. Different paths are currently being intensively explored in the quest for a bulk-boundary correspondence that may allow classifying the topological phases of non-Hermitian lattices~\cite{rudner_survival_2016,leykam_edge_2017,zhou_periodic_2018,ge_topological_2019,brzezicki_hidden_2019}. These forking paths arise because of the many possible options, a very first one is the use of the Hamiltonian or Green's functions as a starting point. In any of the two cases, one might use the initial non-Hermitian single-particle Hamiltonian or a modified one. More flavors are added by the different possible definitions of a gap and the fact that symmetries are enriched by the lifting of the Hermiticity constraint. In the following we mention only a few of the related works.

Early attempts to complete this program include Ref.~\cite{shen_topological_2018}, where the \textit{vorticity}, defined earlier in Section~\ref{section-gap}, appears as a topological invariant. We emphasize that this is at odds with what happens in hermitian topological systems, as the invariant is always computed from the eigenstates not the eigenvalues. One must also note that in Ref.~\cite{shen_topological_2018} one of the underlying assumptions is that the bandstructure is separable, thereby ruling out cases with non-Hermitian degeneracies.

Another proposal for a classification was presented in Ref.~\cite{gong_topological_2018} for the case of the point gaps introduced in Section~\ref{section-gap}. The resulting classification is in analogy with the Hermitian periodic table of topological invariants~\cite{schnyder_classification_2008} in the Altland Zirnbauer symmetry classes~\cite{altland_nonstandard_1997}. This was more recently complemented with a study of cases comprising line-gaps~\cite{kawabata_symmetry_2018}. The general strategy is based on reducing the original problem to a Hermitian one where the classification is known. One of the routes allowing for such a reduction is considering the doubled Hamiltonian~\cite{gong_topological_2018,borgnia_non-hermitian_2019}:

\begin{equation}
{\cal H}_{\mathrm{doubled}}= 
\begin{array}{cc}
% \begin{pmatrix}
  0& {\cal H}^{}  \\
  {\cal H}^{\dagger}& 0  \\
\end{array},
% \end{pmatrix},
 \nonumber
 \end{equation}
which by construction is Hermitian. From the scattering perspective used earlier to interpret defectiveness, one can also interpret the benefit of doubling the Hamiltonian: in this the information which is missing in the non-Hermitian Hamiltonian is restored, thus avoiding the difficulties introduced by defectiveness and the loss of linearly independent eigenvectors.

In most of these works the symmetries of the effective Hamiltonian play a crucial role. As compared with usual Hermitian systems, the symmetries are also much enriched in the non-Hermitian case~\cite{zhou_periodic_2018}. Indeed, symmetries such as particle hole symmetry fork into two since complex conjugation and transposition become distinct~\cite{kawabata_symmetry_2018}, while others get unified, as antiunitary symmetries which are distinct in the Hermitian case and now can be mapped onto each other~\cite{kawabata_topological_2019}. There are also new symmetries which appear for the non-hermitian case, such as pseudo-Hermiticity~\cite{mostafazadeh_pseudo-hermiticity_2002}, while others such as non-hermitian chiral symmetry~\cite{ge_symmetry-protected_2017,rivero_chiral_2019} may remain hidden~\cite{qi_defect_2018}.

In cases where the definition of the topological invariants is attempted directly from the non-Hermitian Hamiltonian, the presence of exceptional points may bring issues such as the non-Hermitian skin effect mentioned in the previous section. Such anomalous localization jeopardizes a direct path connecting a bulk topological invariant to the boundary modes, the sensitivity to the boundary conditions is extreme in this case and the bulk spectrum differs from that of the finite system and, even more, the character of the eigenstates might be disparate (one might have extended states while the other may show only localized states). This motivated other authors to choose non-Bloch invariants~\cite{yao_non-hermitian_2018,yang_non-hermitian_2019}, i.e. invariants based on a finite system rather than bulk states, or invariants defined in real space~\cite{song_non-hermitian_2019}. A similar approach, without reference to Bloch states but rather the system with open boundaries, was presented by Kunst and collaborators~\cite{kunst_biorthogonal_2018}.

Another path is the use of Green's functions as put forward by Zirnstein et al.~\cite{zirnstein_bulk-boundary_2019}. This is motivated by the fact that as one special and useful case of a response function, Green's functions describe observables and could therefore be used as an alternative to the Hamiltonian as the starting point for a topological characterization~\cite{volovik_fractional_1989,volovik_universe_2009}. This has been particularly useful for interacting systems~\cite{gurarie_single-particle_2011}. In the case of non-Hermitian systems, Ref.~\cite{zirnstein_bulk-boundary_2019} focuses on the one-dimensional case where they find that although there might be no correspondence between topological invariants obtained with periodic boundary conditions and
the boundary eigenstates with open boundary conditions, the winding number signals a topological phase transition in the bulk where there is a spatial growth of the Green function.

Borgnia and collaborators~\cite{borgnia_non-hermitian_2019} combined local Green's functions with the doubled Hamiltonian construction to bring together skin effects and bridge the gap for a bulk-boundary correspondence. On a different path, but with the same unifying aim, Kunst and collaborators~\cite{kunst_non-hermitian_2018} put forward a transfer matrix approach. Ref.~\cite{rudner_survival_2016} focused on lattices in one dimension with losses added as on-site terms identifying a winding number.

\section{Connection with Floquet systems}
In this section we explicit a few connections with the physics of the so-called Floquet systems. During the last years there has been an intense activity in the area of driven systems, exploring ways of harnessing light or time-dependent potentials to substantially alter the states of a material in useful ways. A prominent example is inducing chiral edge states in an otherwise normal material by shining a laser on it~\cite{oka_photovoltaic_2009,lindner_floquet_2011,kitagawa_transport_2011}.

Since the prevalent theory for time-periodic systems is Floquet theory~\cite{kohler_driven_2005,moskalets_floquet_2002}, the laser-induced states are referred in the literature as \textit{Floquet edge states}~\cite{oka_photovoltaic_2009,gomez-leon_floquet-bloch_2013,perez-piskunow_floquet_2014} or, more generally, Floquet-Bloch states~\cite{wang_observation_2013} to mean states not necessarily of topological origin. Along this new avenue for topological states, studies have focused on diverse materials and systems including normal insulators,~\cite{lindner_floquet_2011}, graphene~\cite{kibis_metal-insulator_2010,calvo_tuning_2011,zhou_optical_2011,sentef_theory_2015}, other two-dimensional materials~\cite{sie_valley-selective_2015,lopez_photoinduced_2015,huaman_floquet_2019}, Rashba wires~\cite{klinovaja_topological_2016}, ultracold matter~\cite{eckardt_colloquium:_2017}, and topological insulators~\cite{dora_optically_2012,wang_observation_2013,dal_lago_floquet_2015,gonzalez_macroscopic_2016}. Issues such as thermalization~\cite{moessner_equilibration_2017} or even the spontaneous generation of magnetism~\cite{rudner_berryogenesis:_2018} in laser-illuminated systems are some of the latest fronts in this area.

Like non-Hermitian systems, Floquet systems can also be classified as non-equilibrium systems. Indeed, the states can even be unstable and typically one needs active sources or sinks. But besides, this first fact there are many more subtle connecting points. A problem of this type involving a time-dependent term is usually solved in Floquet space, which is the direct product between the usual Hilbert space and the space of time-periodic functions with the period of the Hamiltonian, the latter adds a replica index which can be assimilated to different 'photon' channels~\cite{shirley_solution_1965}. Many observables require projecting over a single replica (like the time-averaged density of states or in a transport calculation where one needs to assume a reference incident channel for scattering), and thus the associated properties correspond to those of the effective Hamiltonian for that portion of Floquet space. This effective Hamiltonian might also be non-Hermitian, but even when not, the mere partition of the space leads to effects similar to those observed in non-hermitian lattices. For example, because of the dissimilar weight on different replicas, there could be Floquet edge states that do not contribute to the transport response and remain silent~\cite{foa_torres_multiterminal_2014}; the coexistence of edge states with a continuum of states belonging to other replicas may introduce a lifetime to those states (see ~\cite{rudner_anomalous_2013,usaj_irradiated_2014,perez-piskunow_hierarchy_2015}), so the spectrum can be considered as effectively complex!

Based on these similarities, many tools used for Floquet systems find a way to non-Hermitian Hamiltonians. This is the case of the doubled Hamiltonian approach mentioned earlier for non-Hermitian systems has been used earlier for Floquet symmetry protected topological phases~\cite{roy_periodic_2017}.

To close this section we note that there are a few studies studying the interplay between non-Hermiticity and driving in  models with gains and losses~\cite{yuce_pt_2015,turker_pt_2018,zhou_non-hermitian_2018,zhou_non-hermitian_2019}, or others with non-reciprocal interactions~\cite{li_floquet-network_2018}. 

\section{Final Remarks}

Besides providing an overview of the literature in this rapid growing field, here we have discussed a few interpretative issues related to concepts like defectiveness, the lack of a full basis of eigenstates for a given non-Hermitian operator. The chosen viewpoint corresponds to scattering theory which provides a starting point to guide our search. The main idea is to think of a non-Hermitian Hamiltonian as an effective description for a parent Hermitian model (on a restricted Hilbert space) plus a boundary condition. The effective system is much like the shadow (in the sense of a projection) of the higher-dimensional parent. Within this picture, defectiveness can be seen as the existence of eigenstates of the parent Hamiltonian which are incompatible with the boundary condition.

The quest for new phenomena shows now many promising possibilities, like the counterpart of higher-order topological states explored by Edvardsson and others~\cite{edvardsson_non-hermitian_2019,liu_second-order_2019,ezawa_electric-circuit_2018}, or the use of non-Hermiticity to model the finite lifetime introduced by electron-electron or electron-phonon interactions~\cite{kozii_non-hermitian_2017} and the search for gapless topological phases in the non-Hermitian arena~\cite{budich_symmetry-protected_2019,carlstrom_knotted_2019,bergholtz_non-hermitian_2019,yang_visualizing_2019}.

From the experimental side, there has been many more experiments aimed at exceptional points in small systems (see discussion in Section 5) than in lattices as discussed here. One noteworthy exception is the experimental observation of bulk Fermi arcs originated from radiative losses in photonic crystal slabs \cite{zhou_observation_2018}. The tipping point is expected to come when the loop closes by feeding new experiments, confirming or refuting existing theory. These are likely to come first in photonics and acoustics or perhaps something that our dear reader has just unveiled. In any case, as the writer said, ``\textit{may your trails be crooked, winding, lonesome, dangerous, leading to the most amazing view}".\footnote{Quote from Edward Abbey's Desert Solitaire, New York, McGraw-Hill, 1968.}

\section*{Acknowledgments}

LEFFT acknowledges support from FondeCyT (Chile) under grant number 1170917. We thank useful discussions with V\'ictor Manuel Mart\'inez Alvarez, Esteban Rodr\'iguez Mena, Alvaro Nu\~nez, Eric Su\'arez Morell, Gonzalo Usaj, Lucila Peralta Gavensky, Carlos Balseiro, Thomas Stegmann, Hee Chul Park, Jung-Wan Ryu, Li Ge and Barbara Dietz. 
%\end{acknowledgments}

\end{document}